\documentclass[twocolumn]{revtex4-1} 
\usepackage{amsmath}
\usepackage{amssymb}
\usepackage{minibox}
\usepackage{microtype}
\usepackage{etoolbox}
\usepackage{bm}
\usepackage[applemac]{inputenc}
\usepackage[usenames,dvipsnames,svgnames,table]{xcolor}
\usepackage{graphicx}

\begingroup
\renewcommand*{\arraystretch}{1.5}
\endgroup
\makeatletter
\renewcommand*\env@matrix[1][\arraystretch]{%
  \edef\arraystretch{#1}%
  \hskip -\arraycolsep
  \let\@ifnextchar\new@ifnextchar
  \array{*\c@MaxMatrixCols c}}
\makeatother

\makeatletter
\apptocmd{\thebibliography}{\global\c@NAT@ctr 10\relax}{}{}
\makeatother

\begin{document}

\title{Josephson effect in fermionic superfluids across the BEC-BCS crossover}

\author{G. Valtolina$^{1,2,3}$}
\author{A. Burchianti$^{1,2}$}
\author{A. Amico$^{1,2}$}
\author{E. Neri$^{1,2}$}
\author{K. Xhani$^{1,2}$}
\author{J. A. Seman$^{1}$}
\altaffiliation[Permanent address:]{
Instituto de F\'isica, Universidad Nacional
Aut\'onoma de M\'exico, Apartado Postal 20-364, 01000 M\'exico
Distrito Federal, Mexico.}
\author{A. Trombettoni$^{4}$}
\author{A. Smerzi$^{1,2,5}$}
\author{M. Zaccanti$^{1,2}$}
\author{M. Inguscio$^{2,6,7}$}
\author{and G. Roati$^{1,2}$}

\affiliation{$^{1}$ INO-CNR Istituto Nazionale di Ottica del CNR, 50019 Sesto Fiorentino, Italy}
\affiliation{$^{2}$ LENS European Laboratory for Nonlinear Spectroscopy, 50019 Sesto Fiorentino, Italy}
\affiliation{$^{3}$ Scuola Normale Superiore, 56126 Pisa, Italy}
\affiliation{$^{4}$ CNR-IOM, Istituto Officina dei Materiali del CNR and SISSA, I-34136 Trieste, Italy}
\affiliation{$^{5}$ QSTAR, Quantum Science and Technology in Arcetri, I-50125 Firenze, Italy}
\affiliation{$^{6}$ Department of Physics and Astronomy, University of Florence, 50019 Sesto Fiorentino,Italy}
\affiliation{$^{7}$ INRIM Istituto Nazionale di Ricerca Metrologica, 10135 Torino, Italy}

\begin{abstract}
We report on the observation of the Josephson effect between two strongly interacting fermionic superfluids coupled through a thin tunneling barrier. We prove that the relative population and phase are canonically conjugate dynamical variables, coherently oscillating throughout the entire crossover from molecular Bose-Einstein condensates (BEC) to Bardeen-Cooper-Schrieffer (BCS) superfluids. We measure the plasma frequency and we extract the Josephson coupling energy, both exhibiting a non-monotonic behavior with a maximum near the crossover regime. We also observe the transition from coherent to dissipative dynamics, which we directly ascribe to the propagation of vortices through the superfluid bulk. Our results highlight the robust nature of resonant superfluids, opening the door to the study of the dynamics of superfluid Fermi systems in the presence of strong correlations and fluctuations.
\end{abstract}

\maketitle

The Josephson effect is a pristine example of a macroscopic quantum phenomenon, disclosing the broken symmetry associated with the superfluid state \cite{Jos}. On a very fundamental level, it allows to pinpoint the most elusive part of the superfluid order parameter, the phase, through a measurable quantity, a particle current \cite{And66}. Furthermore, being based on tunneling processes, Josephson dynamics provides fundamental insights into the microscopic properties of superfluids and their robustness against dissipative phenomena \cite{barone}. Since its discovery, Josephson effect has been demonstrated for a variety of fermionic and bosonic systems \cite{barone,he4,he4a,Cat01,Alb05,Sch05,Lev07,Thy11,Abb13,packardhe3}. However, it has so far eluded observation in BEC-BCS crossover superfluids \cite{Leggett, Zwe11} realized by ultracold Fermi gas mixtures close to a Feshbach resonance \cite{Var07,Novel2}. The interest in these systems is twofold: on the one hand, they encompass the two paradigmatic aspects of superfluidity within a single framework: Bose-Einstein condensation of tightly bound molecules and BCS superfluidity of long-range fermion pairs \cite{Leggett}. Moreover, in the resonant regime where the pair size matches the interparticle spacing, they exhibit universal properties, sharing analogies with other exotic strongly-correlated fermionic superfluids, from cuprate superconductors to nuclear and quark matter \cite{chen,damascelli}.

In this work, we report on the observation of the Josephson effect in ultracold gases of $^6$Li atom pairs across the BEC-BCS crossover. Our Josephson junction consists of two superfluid reservoirs, weakly coupled through a thin tunneling barrier. For all interaction regimes, we detect coherent oscillations of both the pair population imbalance $\Delta N=N_L-N_R$ and the relative phase $\varphi=\varphi_L-\varphi_R$ across the junction, measured {\em in situ} and after time-of-flight expansion respectively. We prove these two observables to be dynamically conjugate \cite{And66}, directly unveiling macroscopic phase coherence in these strongly-correlated fermionic superfluids. We measure the plasma frequency $\omega_J$ of the oscillations around the crossover region, from which we extract the Josephson coupling energy $E_J$ \cite{barone}. 
\begin{figure}[htbp]
\centering
\includegraphics[width=85mm]{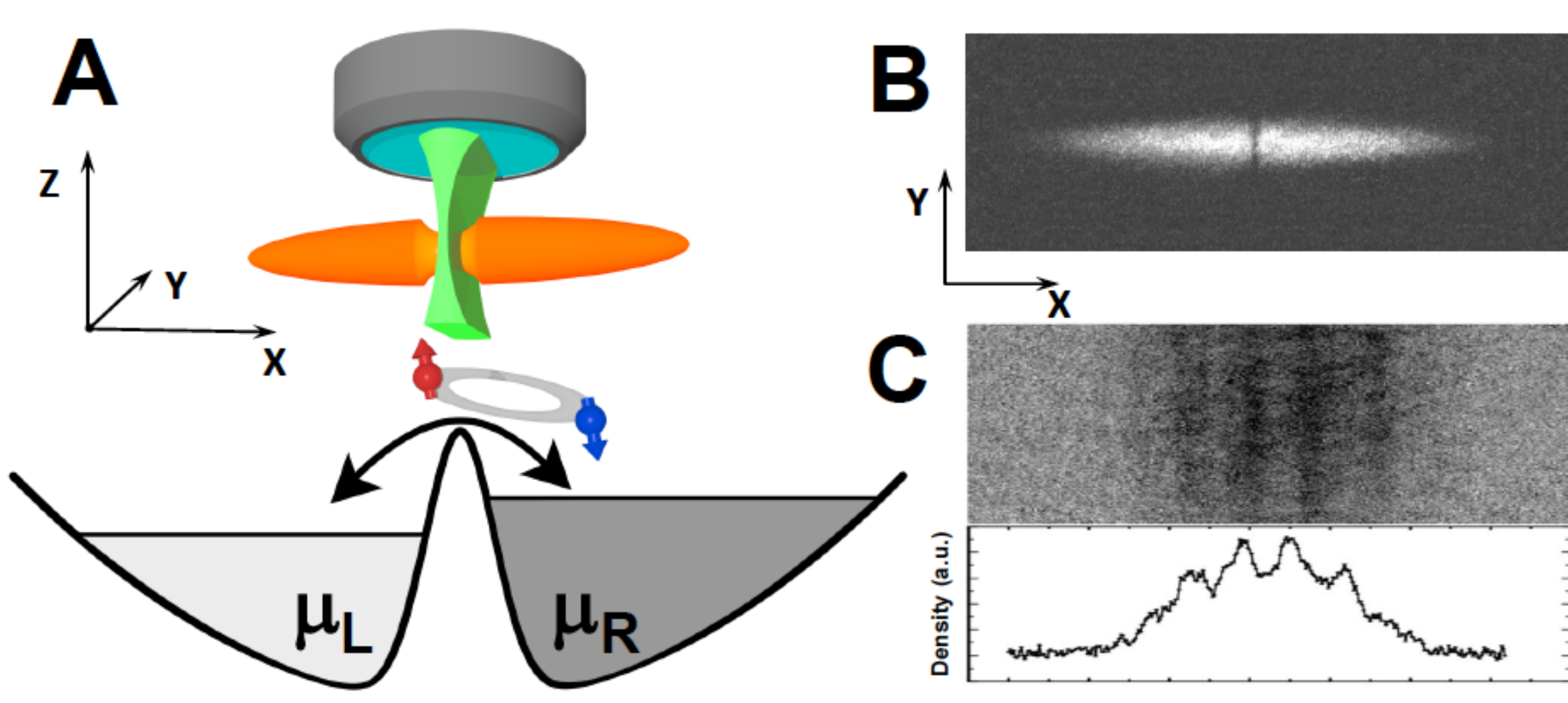}
	\caption{\textbf{A}: Sketch of the experimental apparatus. A Josephson junction is realized by bisecting trapped superfluids of $^6$Li atom pairs with an optical barrier only a few times wider than the correlation length of the system. The dynamics is monitored by recording number imbalance and relative phase between the two reservoirs via \textit{in-situ} (\textbf{B}) and time-of-flight (\textbf{C}) imaging, respectively.}
\label{Fig1}
\end{figure}
Both quantities exhibit a non-monotonic behavior crossing over from the BEC to the BCS regime, reaching a maximum around the resonance region of unitary-limited interactions. For critical parameters of the barrier height or number imbalance, the coherent dynamics turns into a dissipative one. Throughout the whole BEC-BCS crossover, we find that the dissipative flow is triggered by the propagation of vortices in the superfluid bulk, which nucleate within the barrier region via phase slippage processes \cite{Var14,piazza}.
\begin{figure*}[htbp]
\centering
\includegraphics[width=134mm]{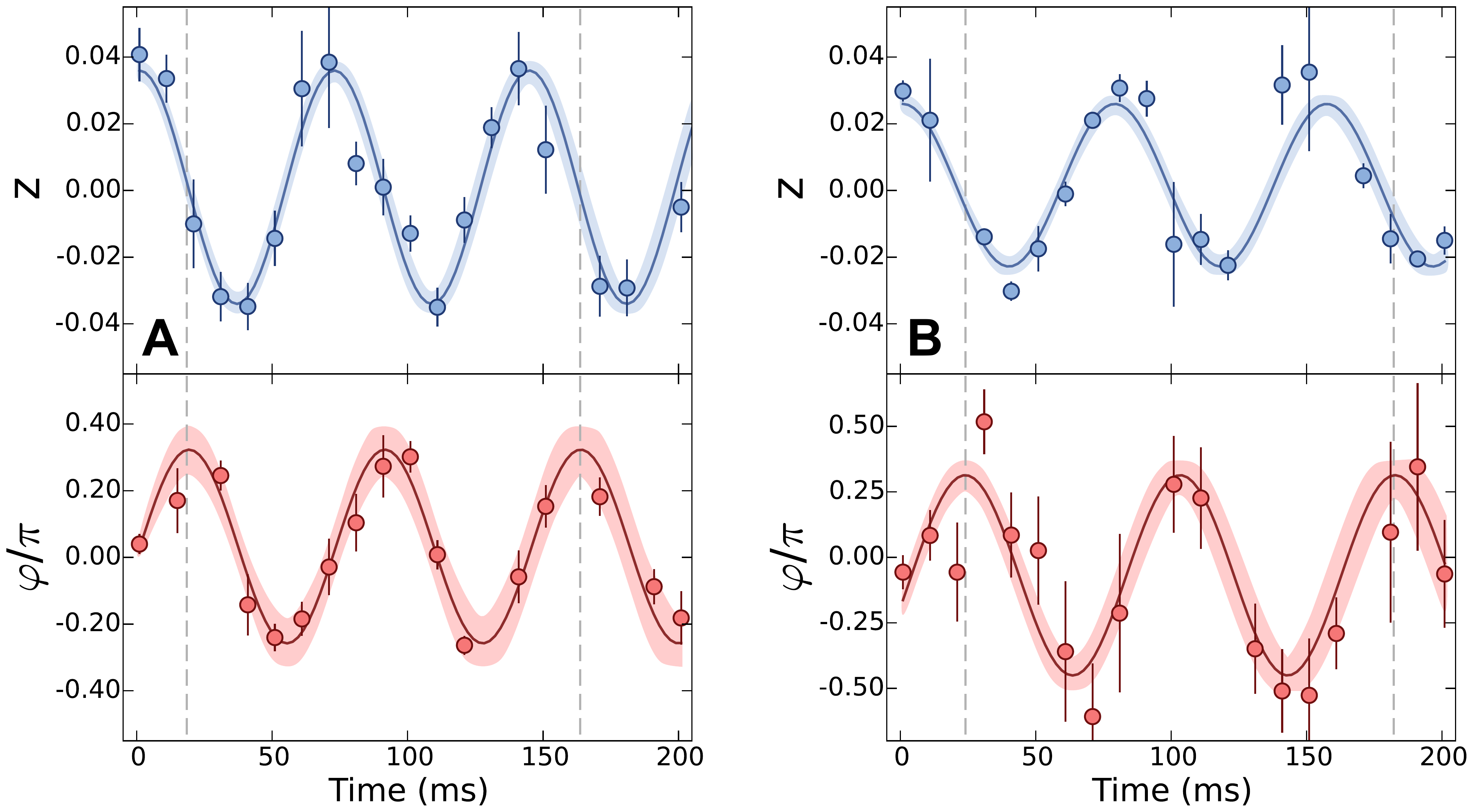}
 	\caption{Josephson oscillations of ultracold Fermi superfluids. \textbf{A}: Time evolution of $z(t)$ (blue) and of $\varphi(t)$ (red) in the BEC limit at $1/k_F a=4.3$ and $V_0/\mu=1.0$. Each data point is the average of at least five independent measurements. Error bars are the corresponding standard deviation. Solid lines are fit to the data with a sinusoidal function. The oscillation frequencies of $z(t)$ and of $\varphi(t)$ obtained from the fit are 13.9(1) Hz and 13.8(2) Hz, respectively. The phase shift between the two oscillations is 1.1(1) $\pi/2$. \textbf{B}: Same as in \textbf{A} but at unitarity for $V_0/\mu=1.1$. Here the phase is detected after a rapid sweep to the BEC limit (see text). The extracted oscillation frequencies are 12.8(1)Hz and 12.6(3) Hz, respectively. The relative phase shift is equal to 1.2(2) $\pi/2$. The shaded regions reflect the fit uncertainties.}
\label{Fig2}
\end{figure*}

In our experiment, we produce fermionic superfluids of about $N=10^5$ atom pairs \cite{Bur14,Materials}, confined in a harmonic potential of frequencies $\omega_x\equiv\omega_0=2\pi\times15$ Hz, $\omega_y=2\pi\times150$ Hz and $\omega_z=2\pi\times170$ Hz. The inter-atomic interaction, parameterized by the $s$-wave scattering length $a$, is finely tuned by exploiting a broad Feshbach resonance at 832 G \cite{Var07,Novel2}. Since strong interactions easily foster the development of dissipative processes \cite{Var14,Sta12}, we engineer a thin optical barrier whose width is only a few times larger than the superfluid coherence length of the system \cite{Lev07}. This is realized by focusing onto the atomic cloud a strongly anisotropic laser beam at 532 nm, blue-detuned with respect to the main optical transition of Lithium atoms. At the trap center, the beam is Gaussian-shaped with 1/$e^2$ beam waist of 2.0(2) $\mu$m and 840(30) $\mu$m along the $x$ and $y$ direction, respectively (see Fig. \ref{Fig1}A). The repulsive barrier is homogeneous along the $y$ and $z$ axes while bisecting the cloud on the weak $x$-axis in two reservoirs. We parametrize the barrier height with the potential peak $V_0$, felt by one atom pair, which is adjusted by controlling the power of the barrier beam. The dynamics of the system is then induced as it follows: initially, the trap center is axially displaced with respect to the barrier position \cite{Materials}, creating an initial non-zero population imbalance $z_0=\Delta N/ N$ between the two reservoirs. Here $N=N_L+N_R$ is the sum of the pair populations of the two wells. A non-adiabatic movement of the trap center back onto the barrier position creates a non zero chemical potential difference $\delta\mu_0$, triggering the superfluid dynamics. Initially, we focus our study on the regime of small excitations, working at the lowest detectable initial imbalance $z_0=0.03(1)$ and for barrier heights $V_0$ exceeding the bulk chemical potential $\mu$, i. e. in the tunneling regime. Here the system Hamiltonian can be written in terms of only two macroscopic dynamically conjugate variables: the relative phase $\varphi$ and the relative population $\Delta N/2$ \cite{And66}. Quite generally, for small oscillations, the Hamiltonian reduces to the sum of two energy terms: $E_J  \varphi^2 /2$ and $E_C/2 (\Delta N/2)^2$. The first one depends on the Josephson tunneling energy $E_J$ and it favors the coherent flow of particles through the junction. The second term represents the charging energy, being $E_C$ the energy cost to add a single pair in one reservoir \cite{And66,Sme97}. Consequently both imbalance and phase undergo harmonic oscillations, out-of-phase by $\pi/2$, at a plasma frequency that is independent from $z_0$ and given by:
\begin{equation}
	\hbar \omega_J=\sqrt{E_C E_J}
\label{eq1}
\end{equation}
We confirm this expectation by studying the evolution of $z(t)$ and $\varphi (t)$ from absorption images, recorded \textit{in-situ} and after time-of-flight expansion respectively, see Fig. \ref{Fig1}. To determine $\varphi(t)$ also in the strongly interacting regime, we perform a $200~\mu s$ fast ramp to the BEC side of the resonance \cite{Materials,Ket06}. This reduces the detrimental effects of collisions during the expansion, that would completely wash out the visibility of the interferogram \cite{Kohstall11}. In Fig. \ref{Fig2} we present an example of the evolution of $z(t)$ and $\varphi (t)$ for a molecular BEC and a unitary Fermi gas, respectively. In both cases, the two quantities oscillate at the same frequency, with a relative phase shift of $\pi/2$, within error bars. Our measurement proves the conjugate nature of phase and number imbalance, representing a direct proof of the macroscopic phase coherence of these systems. Since this trend is confirmed also for a crossover superfluid, whose phase can be inferred only after ramping to the BEC limit, it indicates that the magnetic sweep does not appreciably distort the measurement of $\varphi(t)$.

In order to understand how the Josephson dynamics is influenced by interactions through the BEC-BCS crossover, we extract the Josephson frequency $\omega_J$ by measuring $z(t)$ for a fixed barrier height $V_0=1.2(1) E_F$, at different values of the interaction strength that we parametrize via the dimensionless parameter $1/k_F a$. 
\begin{figure}[htbp]
\centering
\includegraphics[width=75mm]{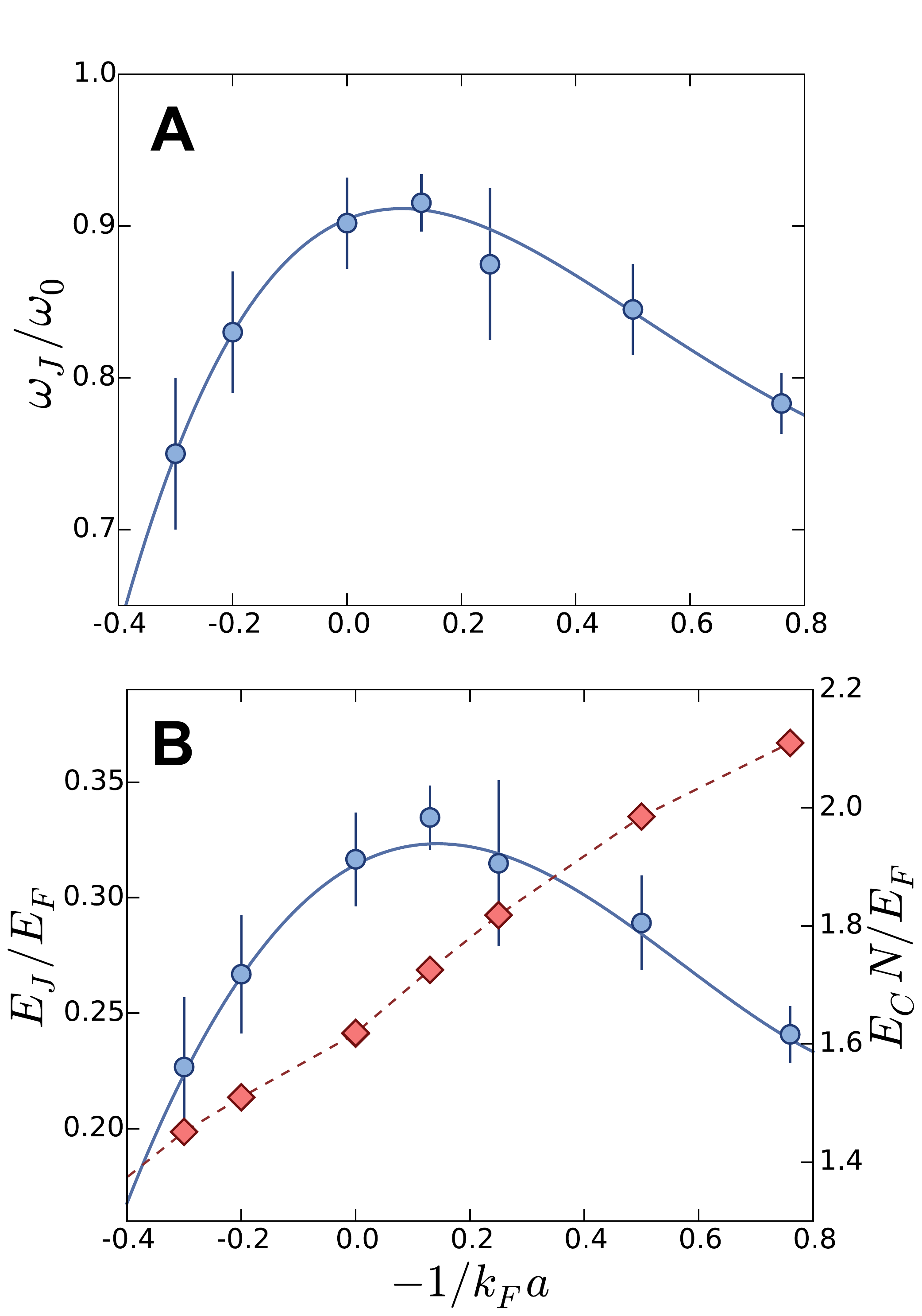}
 \caption{Plasma frequency and Josephson coupling energy through the BEC-BCS crossover. \textbf{A}: $\omega_J$ (blue circles) normalized to the bare trap frequency $\omega_0$ for $V_0/E_F=1.2(1)$ as a function of $1/k_F a$.  Each data point is the average of at least five independent measurements. Error bars are the corresponding standard deviation. \textbf{B}: Calculated charging energy $E_C$ multiplied for the number of pairs N (red solid diamonds) and Josephson coupling energy $E_J$ (blue circles) normalized to the Fermi energy E$_F$ extracted from Eq. (\ref{eq1}). In both panels blue solid lines are a guide to the eye.}
\label{Fig3}
\end{figure}
Here $k_F$ is the Fermi wave-vector associated to a non-interacting gas of $N$ trapped fermions, whose Fermi energy is given by $E_F=\frac{\hbar^2 k_F^2}{2 m}=\hbar \varpi(6 N )^{1/3}$ ($m$ is the mass of one $^6$Li atom and $\varpi=(\omega_x \omega_y \omega_z)^{1/3}$). In Fig. \ref{Fig3}A we present the behavior of $\omega_J$ normalized to $\omega_0$ as a function of $1/k_F a$. Notably, the extracted $\omega_J$ exhibits a non-monotonic evolution across the BEC-BCS crossover, with a maximum around the unitary limit. To gain further insights into this trend, by exploiting Eq. (\ref{eq1}) we derive $E_J$  from the measured values of the plasma frequency $\omega_J$ combined with the computed charging energy.  $E_C$ is related to the inverse of the system compressibility, and it has been derived from an extended Thomas-Fermi model, based on a generalized Gross-Pitaevskii equation for the pairs wavefunction that accounts for the correct chemical potential obtained by Monte Carlo calculations across the entire crossover \cite{Materials}. In Fig. \ref{Fig3}B we show $E_J$ and $N\times E_C$ normalized to the Fermi energy $E_F$ as a function of $1/k_F a$. While the charging energy increases monotonically moving from the BEC to the BCS regime, the Josephson coupling energy reflects the behavior of $\omega_J$, reaching a maximum close to unitarity. Interestingly, our observation qualitatively reproduces the trend for the maximum Josephson current derived from numerical approaches \cite{Spu07,Zou14} which explicitly account for the composite nature of the superfluid pairs. For $V_0\ll\mu$, this behavior can be ascribed \cite{Spu07} to the competition of two different critical velocities at the crossover region, set by sound and pair-breaking excitations that are predominant on the BEC and BCS side respectively \cite{ketterle}.  In turn, our observation in the tunneling regime $V_0>\mu$ can be qualitatively understood by noting that in both the BEC and BCS limits one expects
\begin{equation}
	E_J \sim{\cal K} N_0.
\label{eq-cond}
\end{equation}
$N_0$ is the total number of condensed pairs, while ${\cal K}$ is the tunneling term that depends on the chemical potential and the barrier properties \cite{Sme97,zwerger}. In the deep BEC limit, almost all pairs are condensed, $N_0\simeq N$ and $E_J\sim{\cal K}N$. In the BCS limit $N_0/N \propto \Delta/E_F$, where $\Delta$  is the superfluid gap, Eq. (\ref{eq-cond}) reproducing the Ambegaokar-Baratoff formula at $T=0$ \cite{barone}. Moving from the BEC to the unitary limit the increase of $E_J$ reported in Fig. \ref{Fig3}B is associated with the growth of ${\cal K}$. Indeed, in this regime $N_0$ decreases only slightly \cite{Astra}, while the increase of the interactions makes $\mu$, hence  ${\cal K}$, progressively larger. In contrast, towards the BCS limit, $\mu$ does not vary much with $k_Fa$, while $N_0$ is strongly decreased, causing a net reduction of $E_J$. Even if a microscopic derivation of the dependence of $E_J$ upon $N_0$ is presently missing in the strongly-interacting regime, the previous arguments suggest that a maximum of $E_J$ should occur close to the unitary limit, in agreement with the experiment.
\begin{figure*}[htbp]
\centering
\includegraphics[width=140mm]{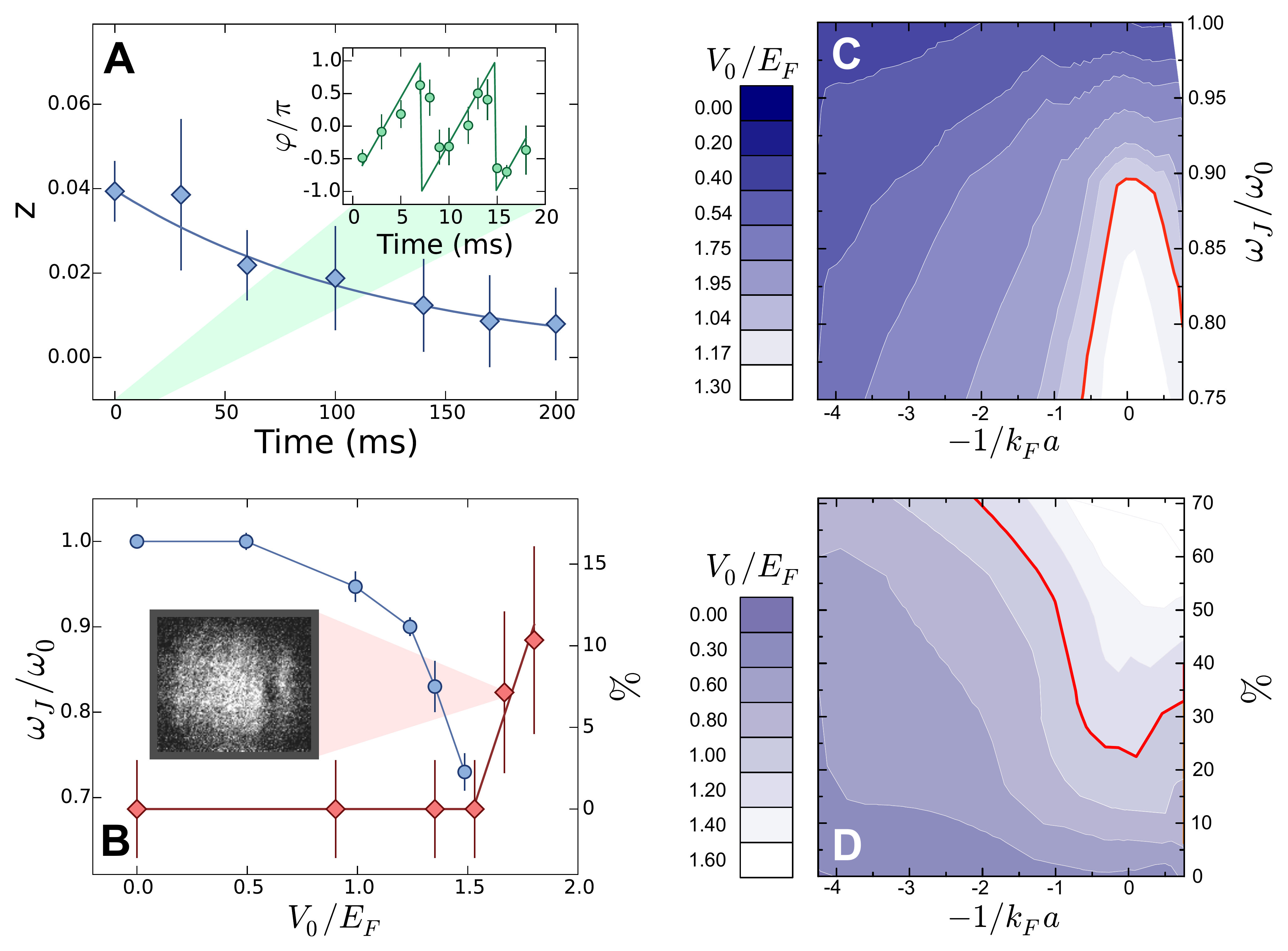}
 \caption{Onset of dissipative dynamics. \textbf{A}: Time evolution of $z(t)$ at resonance for $ V_0/E_F=$ 2 and $z_0=0.04(1)$ (blue diamonds). Solid line is an exponential fit to the data. Inset: Time evolution of $\varphi(t)$ for the same initial parameters (green circles). The green solid line is a fit to a sawtooth function. Error bars are one standard deviation of five independent measurements. \textbf{B}: Vortex occurrence probability evaluated over a collection of 40 independent time-of-flight images (red diamonds, right axis), and normalized plasma frequency $\omega_J/\omega_0$ (blue circles, left axis), as a function of $V_0/E_F$ at $1/k_F a=0$, for $z_0=0.04(1)$. Error bars for the vortex occurrence are obtained by the Wilson score interval. Error bars for $\omega_J/\omega_0$ represent the fit uncertainties. Inset: Typical image of one vortex, recorded after 10 ms time-of-flight expansion. \textbf{C}: Contour plot of $\omega_J/\omega_0$ versus $1/k_F a$ for different $V_0/E_F$ values indicated by the color code on the left. \textbf{D}: Same as in C for the probability of vortex occurrence. To speed up the data acquisition, $z_0=0.12(2)$ and the statistics was performed over 20 images. In panels C and D the red line highlights the barrier value $V_0/E_F=1.2$.}
\label{Fig4}
\end{figure*}

So far, we have focused our studies on the regime of small excitations, characterizing the Josephson oscillations throughout the BEC-BCS crossover. Now, we investigate the system evolution once the dynamics is triggered by larger initial excitations. In our setup, this can be easily done either by increasing the initial imbalance $z_0$ at fixed $V_0$, or vice versa. In this case, we observe a completely different behavior, where the transport through the junction turns from coherent to dissipative. As an example, in Fig. \ref{Fig4} A we present the evolution of $z(t)$ for $z_0=0.04(1)$ and $V_0/E_F=2$, at $1/k_Fa=0$. This dynamics is in striking contrast with the one previously reported in Fig. \ref{Fig2}: the system is now characterized by a dissipative flow, which irreversibly tends to equilibrate the two reservoirs, similarly to what observed in Ref. \cite{Sta12} for crossover superfluids coupled through a mesoscopic channel. In this regime, it is interesting to study also the behavior of $\varphi(t)$. A priori, phase coherence between the two superfluids could be completely scrambled by the presence of dissipative processes. In turn, the visibility of the observed interference pattern remains high, and we are able to trace the initial evolution of $\varphi(t)$, see inset of Fig. \ref{Fig4} A. Also the dynamics of this observable significantly differs from the one in Fig. \ref{Fig2}: $\varphi(t)$ grows now linearly in time, at a rate much faster than $\omega_0$ and comparable with the initial chemical potential imbalance $\delta \mu_0$. This behavior can be explained by noticing that, once the initial charging energy $E_Cz_0^2N^2$ exceeds the Josephson coupling $E_J$, the phase increases as $\hbar \varphi(t) \sim E_C z_0 N t$ \cite{Sme97}. 
In the experiment, after an evolution time on the order of the axial trap period, despite the contrast of the interferograms remains constantly high in each image, we observe a significant increase of the shot-to-shot fluctuations that impedes to further follow a clear trend of $\varphi(t)$. In order to understand how a slow resistive flow can coexist with the phase coherence between the two reservoirs, we investigate the microscopic origin of dissipation in our system. The high visibility of the interference pattern, detected at all evolution times, rules out pair-breaking processes, characterizing weakly interacting BCS superfluids, as the origin of the dissipative flow. Rather, our observation recalls the phenomenology typical of Helium systems, where resistive dynamics is established by phase-slippage processes and vortex nucleation. In fact, the entrance into this "running phase" regime is expected to be accompanied by the nucleation of a vortex within the link region \cite{Var14}. 
A complete phase slip $\varphi=2\pi$ occurs when the vortex annihilates within the barrier region. This gives rise to the macroscopic quantum self trapping of the relative population \cite{Materials,Sme97}. 
However, depending on the system configuration, and for $V_0\sim \mu$, such a topological defect may escape the low-density region \cite{And66, Var14,piazza} entering the superfluid bulk, before annihilation. The propagation of the vortex through the superfluid bulk acts as a dissipative channel that gives rise to a resistive flow that leads to an exponential decay of $z(t)$. This mechanism can indeed occur in our crossover superfluids: the three-dimensional character of our junction, combined with the coupling to the transverse modes favored by the strong inter-particle interactions, may facilitate the leakage of vortices from the barrier region \cite{Campbell13}. Indeed, by performing a statistical study over several time-of-flight images recorded after some time evolution in the trap \cite{Materials}, we detect with non-zero probability the presence of topological defects, which appear as density depletions in the expanded clouds, see inset in Fig. \ref{Fig4} B. By measuring their oscillation period in the trap after switching off the barrier we identify them as solitonic vortices \cite{Materials,Ku}. The intimate connection between the breakdown of the Josephson oscillations and the appearance of vortices is further confirmed by the study presented in Fig. \ref{Fig4} B. 
Here, we show the behavior of the Josephson frequency $\omega_J$ at unitarity as a function of $V_0$, together with the occurrence of defects collected for each $V_0$ value over a statistical ensemble of 40 images. No vortices are detected until coherent oscillations are observed ($V_0/E_F<1.5$), topological defects showing up only after the stop of coherent oscillations. Interestingly, the interconnection between the quench of the coherent dynamics and the vortex nucleation is not peculiar of the unitary point, but it extends over the whole BEC-BCS crossover region. This can be clearly observed by comparing Fig. \ref{Fig4} C and D, where the measured $\omega_J$ is contrasted to the vortex occurrence probability, as a function of $V_0/E_F$ and $1/k_Fa$. One can see how the trend of the first observable perfectly mirrors the behavior of the second one for all interaction regimes. In particular, Fig. \ref{Fig4} D highlights once more the particular robustness of the crossover superfluid, which shows a reluctance to the formation of topological defects while keeping the highest Josephson frequency. We also note that the results of our measurements differ from the ones reported when investigating the limit of vanishingly low barriers $V_0\ll\mu$, where phononic excitations and pair-breaking effects, rather than vortices, cause the breakdown of superfluidity in the BEC and BCS side respectively \cite{ketterle}. 

In conclusion, we have observed coherent Josephson dynamics between two weakly-coupled superfluids of $^6$Li atoms across the whole BEC-BCS crossover. The behavior of the Josephson coupling energy $E_J$, maximum nearby unitarity, together with the response to excitations, minimum at the resonance, provides further evidence of the high-T$_C$ nature of these fermionic superfluids also in the coherent tunneling regime. Our work paves the way to the study of the interplay between elementary and topological excitations in the dissipative dynamics obtained by varying the height and width of the interwell barrier and to the measurement of the superfluid gap in close analogy with tunneling experiments in superconductors \cite{barone,damascelli}. Moreover, extending our studies of the tunneling dynamics above the condensation temperature $T_C$ may give insights on the role of phase fluctuations in the regime where preformed, non condensed pairs appear in the system \cite{chen}.\\

\textbf{Acknowledgments:} We acknowledge inspiring discussions with F. Dalfovo, A. Recati  and W. Zwerger. We thank C. Fort, A. Trenkwalder, A. Morales and T. Macr\'i for collaboration at the initial stage of this work.  Special acknowledgments to the LENS Quantum Gases group. This work was supported under the ERC Grant No.307032 QuFerm2D.

\renewcommand{\thefigure}{S\arabic{figure}}
 \setcounter{figure}{0}
\renewcommand{\theequation}{S.\arabic{equation}}
 \setcounter{equation}{0}
 \renewcommand{\thesection}{S.\Roman{section}}
\setcounter{section}{0}
\renewcommand{\thetable}{S\arabic{table}}
 \setcounter{table}{0}

\onecolumngrid

\newpage


\begin{center}
{\bf \large Supplementary Materials for\\
``Josephson effect in fermionic superfluids across the BEC-BCS crossover''}

\bigskip

G. Valtolina, A. Burchianti, A. Amico, E. Neri, K. Xhani,\\J. A. Seman, A. Trombettoni,
A. Smerzi, M. Zaccanti, M. Inguscio, and G. Roati

\end{center}

\bigskip
\twocolumngrid

\section{Sample preparation}
\label{sec:1}

Fermi superfluids are prepared by evaporating a two-component mixture of the lowest hyperfine states of $^6$Li in an optical dipole trap. We employ the $\left|F=1/2,m_F=\pm1/2\right\rangle$ states (unless otherwise stated), labeled as $\left|1\right\rangle$ and $\left|2\right\rangle$. Following the procedure described in Ref.~\cite{Bur14a}, the atomic sample evaporated at a magnetic field of $832$~G, on top of a broad Feshbach resonance, finally resulting in the production of a superfluid cloud of 1.0(1)$\times$10$^5$ atoms per spin state. At the end of the evaporation, the magnetic field is adiabatically ramped to the desired value, allowing fine tuning of the atomic scattering length $a$, evaluated using the magnetic field dependence $a({B})$ reported in \cite{Zurn13}. The trap is realized by two infrared laser beams, crossing with an angle of $14^\circ$, in the $xy$-plane. The primary beam has a wavelength of $1064$~nm and it is focused on the atomic sample on a waist of 45~$\mu$m. The secondary beam has a wavelength of $1070$~nm and waists on the atoms of 100 and 45~$\mu$m. The final harmonic potential is characterized by radial trapping frequencies of about $150$ and $170$~Hz along the y and z axes respectively and an axial frequency of about $15$~Hz. The magnetic field curvature of the Feshbach coils provides additional confining (anti-confining) potential along the x and y (z) directions, corresponding to trap (anti-trap) frequencies of $\omega_{x,y}\sim2\pi\times8$~Hz and $\omega_{z}= \sqrt{2}\omega_{x,y}$~Hz at 832 G. These contributions sum in quadrature to the main optical trap frequencies, causing a variation of the overall trap frequencies of about 10$\%$ when spanning from the BEC to the BCS limit.

\subsection{Creation of the tunneling barrier}
We produce the tunneling barrier by focusing on the atoms an anisotropic laser beam at $532$ nm, by using a single aspheric lens (NA$\sim 0.6$). This configuration produces a repulsive sheet of light which bisects the trapped cloud. The repulsive barrier is first characterized on a replica of the optical system implemented in the experiment. The measured beam waists, detected with a CCD camera, in the focus position are 1.9(1)~$\mu$m  and  840(30)~$\mu$m, along the $x$ and $y$-axis, respectively. To extrapolate the barrier width in the final set-up, we study the atomic density profiles in the BEC regime, exploiting the Thomas-Fermi approximation $n(\textbf{r})$=$\text{max}((\mu-V(\textbf{r}))/g,0)$, where $\mu$ is the chemical potential, $V(\textbf{r})$ is the double well-potential and $g$ the interaction coefficient.
\begin{figure}[t]
\centering
\includegraphics[width=80mm]{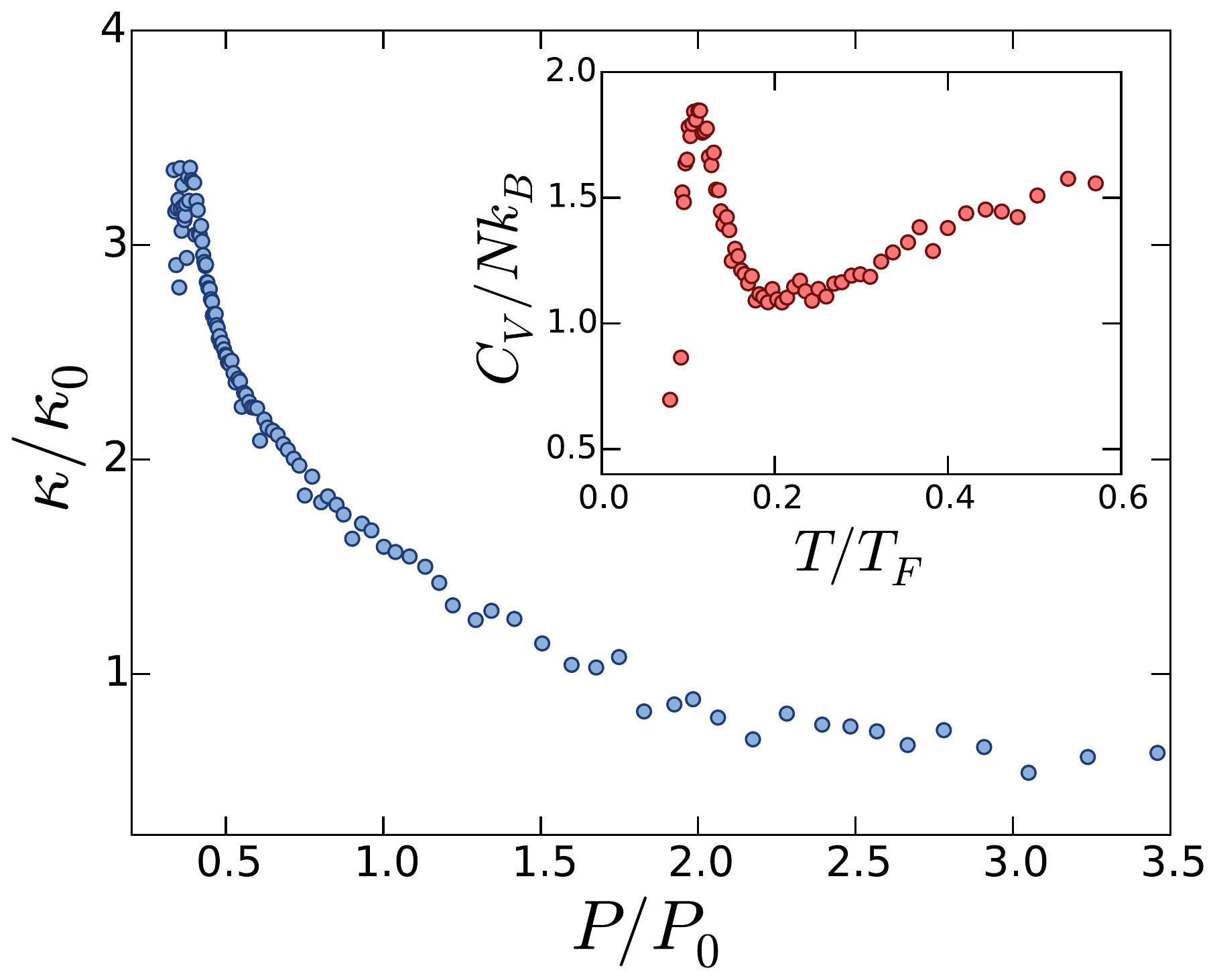}
\caption{Equation of state of the unitary Fermi Gas. \textbf{Main}: Normalized compressibility (blue circles) $\kappa/\kappa_{0}$ versus normalized pressure $P/P_{0}$. $\kappa$ and $P$ are scaled by the respective quantities $\kappa_{0}$ and $P_{0}$ at the same local density for a non-interacting Fermi gas. \textbf{Inset}: Specific heat per particle (red circles) $C_{V}/Nk_{B}$ as a function of the reduced temperature $T/T_{F}$.}
\label{FIG0S}
\end{figure}
In-situ absorption images of the atomic density distribution in the $xy$-plane are taken by progressively decreasing the intensity level of the probe beam. This reduces the effect of atom diffusion during the imaging pulse, which can distort the profile of the barrier beam carved in the atomic cloud. By extrapolating the barrier size to zero intensity we find $w_{x}$=$2.0(2)$~$\mu$m. Numerical simulations of the dynamics in the BEC limit (see section \ref{sec:7}), in which the beam waist enters as a fixed parameter, further confirm the measured value of $w_{x}$. 

\subsection{Control of the initial population imbalance and of the initial velocity}
The position of the $1070$~nm secondary beam, relative to the main one, is finely adjusted by changing the radio-frequency of the acousto-optic modulator controlling its optical power. The change in the RF signal results in a shift of position of the beam focus. The $1070$~nm beam mainly provides the trap axial confinement and its movement results only in an axial shift of the trap minimum. The barrier potential is then held fixed at a certain place which we identify as the final position of the trap minimum. Evaporation is performed in a trap displaced with respect to the barrier position in order to create a population imbalance among the left and right sides. The dynamics is triggered by non-adiabatically shifting the trap minimum over the barrier position. It is worth noticing that for the typical initial imbalances employed in this work, $z_0=0.03(1)$, the maximum atomic velocity relative to the barrier is of the order of $v=0.03(1) v_\text{F}$, where $v_\text{F}$ is the Fermi velocity. For any interaction regime this value lies well below the recently measured Landau critical velocity \cite{Wei15} for breaking the superfluid. 

\section{Thermometry and Equation of State}

A first indication about the temperature at the end of evaporation is obtained by measuring the condensed fraction in the BEC regime. Without the barrier, we estimate a condensed fraction always above $90\%$ of the total pair number. The condensed
fraction is determined by time of flight images recored after a Feshbach magnetic field sweep towards the BEC limit ($690$~G), by fitting the momentum distribution with a 2D bimodal function. Evaporation with the barrier on at any target height does not result in a change of the condensed fraction. Moreover, the condensed fraction does not vary appreciably over the experimentally accessed evolution times, for all the different tunneling regimes.
We further confirm the presence of a superfluid state at unitarity by determining the equation of state of the gas, following the procedure developed in \cite{Ku12}. In particular, we observe a peak in both the compressibility as a function of the pressure and in the specific heat, as a function of T/$T_F$, at a critical temperature $T_c=0.15(3)T_F$, where $T_F=E_F/k_{B}$ ((being $k_{B}$ the the Boltzmann constant), as shown in Fig.~\ref{FIG0S} and in the inset, respectively. Our measured value of $T_c$ is in fair agreement with the one reported in \cite{Ku12}. We can extract the temperature of the gas by employing the virial expansion on the data. In this way, we estimate a degeneracy parameter at the trap center of $T/T_F=0.07(2)$. This gives a ratio $T/T_c$ below 0.6, which ensures a superfluid fraction around unity for the resonantly interacting Fermi gas \cite{Sid13}.

\section{Phase measurement}
\label{sec:4}

We study the evolution of the relative phase by detecting the interference fringes arising from the two expanding clouds after a time of flight of typically 15 ms. The phase is extracted by fitting the resulting interferogram by a 2D Gaussian, modulated by a cosine function, of the form $n(x,y)=A e^{-x^2/w_x^2}e^{-y^2/w_y^2}\times(1+Bcos(kx+\varphi))$. In the BEC side of the resonance, we can extract the relative phase at the same magnetic field at which the dynamics of the imbalance is observed. For the acquisition of the data in the unitary limit and in the BCS side, we employ a balanced mixture of the $|1\rangle-|3\rangle$ states, where $\left|3\right\rangle$ corresponds to the $\left|F=3/2,m_F=-3/2\right\rangle$ state at low magnetic fields. In this case, the sample is produced by performing the evaporation at $690$~G, on top of the $|1\rangle-|3\rangle$ Feshbach resonance. This resonance has a width significantly smaller than the one at $832$~G. This facilitates the fast magnetic field sweep from the unitary to the BEC regime for increasing interference patterns visibility \cite{Ket06a}. We have also checked the $|1\rangle-|3\rangle$ population imbalance to evolve with the expected coupled dynamics with respect to the phase, and in accordance with the measurements taken on the $|1\rangle-|2\rangle$ mixture in the same initial condition.
We can also observe the transition towards completely disconnected superfluids when $V_0\geq4\mu$.The entrance in this disconnected regime is signaled by an increase in the variance $\Delta\varphi^2$ that approaches to $2\pi$.
\begin{figure}[t]
\centering
\includegraphics[width=80mm]{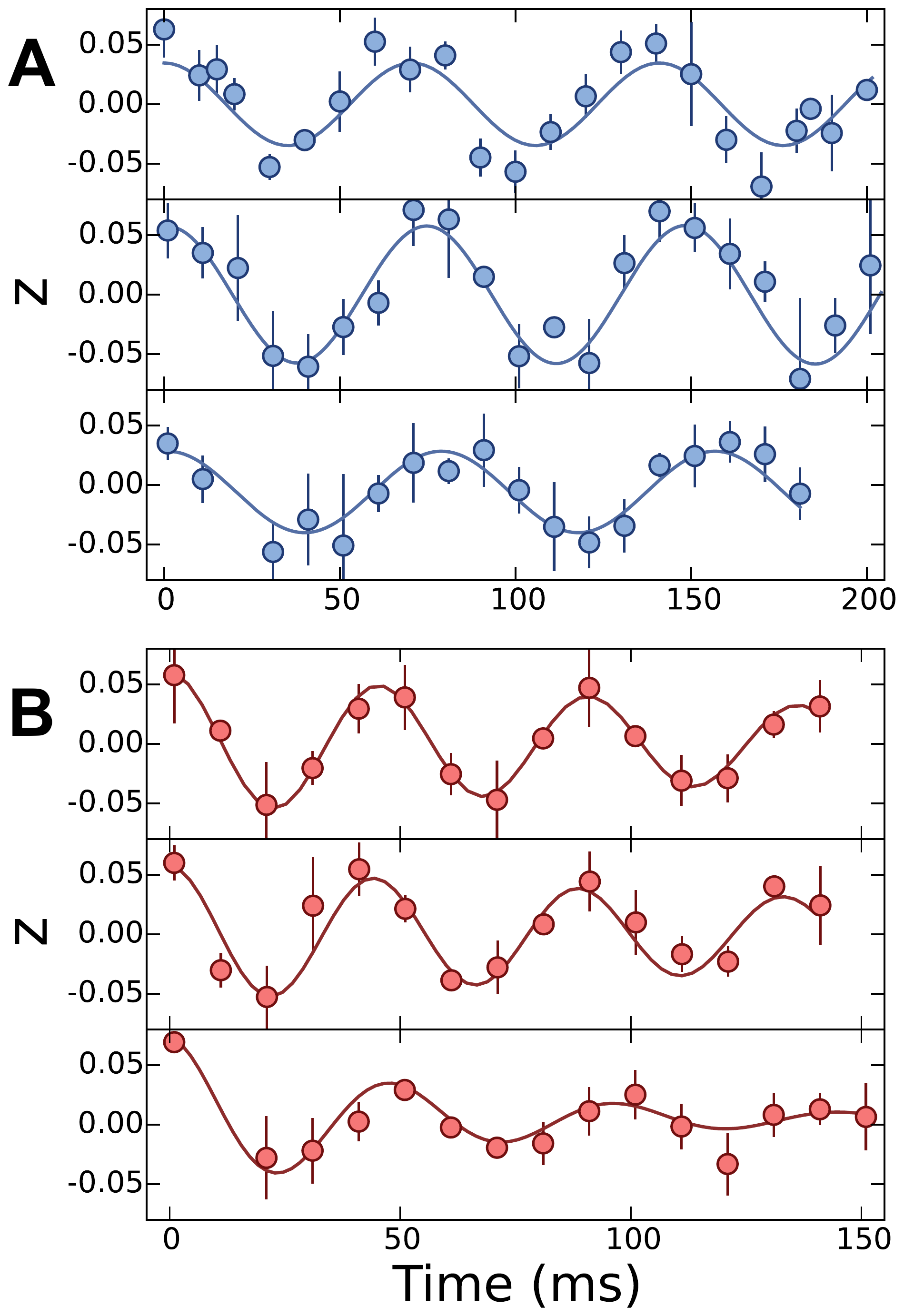}
\caption{Superfluid versus normal tunneling.  \textbf{A:} The superfluid oscillations (blue circles) are shown for different barrier heights. The values of $V_0/\epsilon$ are: $0.25(2)$ (top), $1.50(3)$ (center), $1.75(4)$ (bottom). \textbf{B:} The normal gas oscillations (red circles) are shown for different barrier heights. The values of $V_0/\epsilon$ are $0.40(2)$ (top), $0.60(2)$ (center), $0.81(3)$ (bottom). In all the plots, the error bars are one standard deviation of five independent measurements.}
\label{FIG1S}
\end{figure}

\section{Superfluid versus Normal tunneling dynamics}

\begin{figure}[htpb]
\centering
\includegraphics[width=80mm]{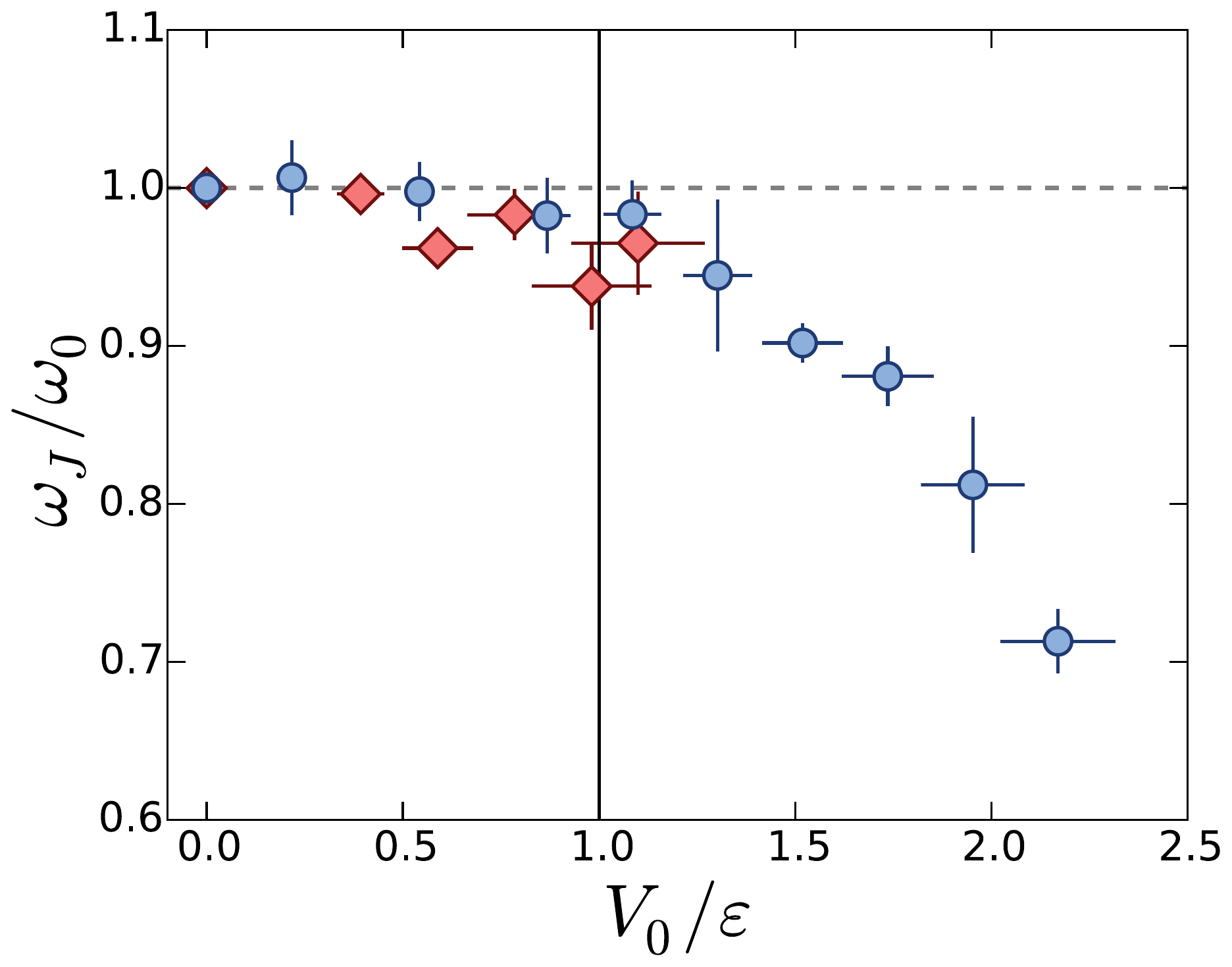}
\caption{Superfluid and normal gas oscillations versus the barrier height.  Oscillation frequency, in units of the trap frequency $\omega_{0}$, as a function of $V_0/\epsilon$ for the superfluid (blue circles) and the normal gas (red diamonds). In all the plots, the vertical error bars are one standard deviation of five independent measurements, while the horizontal error bars take into account the experimental uncertainties of the atoms number and of the barrier width.}
\label{FIG1S_B}
\end{figure}

In this section we compare the dynamics of a superfluid molecular BEC ($1/k_F a=4.6$) with the one of a spin polarized Fermi gas at $T/T_F\sim 0.1$. The latter is produced by evaporating a balanced $|1\rangle-|3\rangle$ mixture at $300$~G \cite{Bur14}. The scattering length is brought to zero by setting the Feshbach field around $570$~G, in the vicinity of the zero-crossing of the $|1\rangle-|3\rangle$ Feshbach resonance. The experimental results are shown in Fig.~\ref{FIG1S} and Fig.~\ref{FIG1S_B}. Here, it is convenient to rescale the barrier height to the mean energy per particle $\epsilon$ instead than the chemical potential $\mu$. In the BEC limit $\epsilon=\frac{5}{7}\mu$, while for the normal Fermi gas $\epsilon=\frac{3}{4}E_F$. As already discussed in the main text, the superfluid undergoes coherent undamped oscillations also for barrier heights larger than $\epsilon$ (see Fig.~\ref{FIG1S}~A). For the normal, non interacting Fermi gas this is not the case. 
By increasing the barrier height, the dynamics of the system is characterized by an increasing damping of the oscillations, see Fig.~\ref{FIG1S}~B). For $V_0/\epsilon>1$, no oscillations are detected, the motion of the normal gas becoming overdamped in this regime. Moreover, while the superfluid systems exhibits a strong renormalization of the frequency as $V_0$ is increased, the normal gas oscillates always at the bare trapping frequency $\omega_0$, as shown in Fig.~\ref{FIG1S_B}.

\section{Extended Thomas-Fermi model}
\label{sec:7}
In this section we briefly discuss the theoretical framework we use both for the determination of $E_C$ in Fig. 3B of the main text, and for the numerical simulations of the system dynamics. In particular we make use of the so called  extended Thomas-Fermi model (ETFM) \cite{Man05,Salasnich,McN14,Bul14}. In the ETFM approach, the 
wavefunction $\Psi$ obeys the equation
\begin{equation}
i \hbar \frac{\partial \Psi}{\partial t}=-\frac{\hbar^2}{4m} 
\nabla^2 \Psi+2V\Psi+
2F( 2 \vert \Psi \vert^2) \Psi\,,
\label{ETFM_eq}
\end{equation}
where $\vert \Psi(\vec{r,t}) \vert^2$ is the pair density normalized to the number of pairs $N$ (the total number of particles being $2N$), $V(\vec{r})$ is the external potential felt by the atoms (having mass $m$). The nonlinear term in Eq. (\ref{ETFM_eq}) is defined as $F(n)=\frac{\partial{\cal E}}{\partial n}$, where 
$n=2\vert \Psi \vert^2$ is the total fermion density and ${\cal E}$ is the energy per particle, which has been determined across the BEC-BCS crossover by Monte Carlo calculations \cite{Gan11}. The choice of the nonlinear term $F$ guarantees that the chemical potential of the bulk system, $\mu$, is correct. By writing $2F( 2 \vert \Psi \vert^2) \equiv f(\vert \Psi \vert^2)$ one has in the BEC side $f(\vert \Psi \vert^2)=\frac{4 \pi \hbar^2 a_M}{M} \vert \Psi \vert^2$ (where $M=2m$ is the mass of the pair and $a_M$=0.6$a$ is the molecular scattering length) and at unitarity $f(\vert \Psi \vert^2)=2\xi \frac{\hbar^2}{m} (6\pi^2)^{2/3} \vert \Psi \vert^{4/3}$ (where $\xi$ is the Bertsch parameter: $\xi\approx 0.370$).  The potential $2V \equiv V_p$ acting on the pairs has the form 
\begin{equation}
V_p(\vec{r})=\frac{1}{2} M \left(\omega_x^2 x^2+\omega_y^2 y^2+\omega_z^2 z^2\right) +V_0 e^{-2x^2/w^2}\,
\label{Vp_eq}
\end{equation}
where $\omega_i$ ($i\equiv x,y,z$) are the trap frequencies and $V_0$ and $w$ are the barrier height and width, respectively. In all numerical simulations  the values of $w$, $V_{0}$, $\omega_i$, $N$ and $a$ are fixed to the experimental ones. The initial population imbalance is created, in the framework of the ETFM, as it follows: the center along $x$ parabolic term in Eq.~\ref{Vp_eq} is suddenly displaced to an initial value $x_0$ and then put back in $x=0$. We considered different realistic times for this process and verified that the effect on the tunneling dynamics is negligible. 
From the solution $\Psi\left(\vec{r},t\right)$ one can extract the number of pairs 
$N_L(t),N_R$(t) in the left and right well, respectively, and consequently $z(t)$. 

\begin{figure}[h!]
\includegraphics[width=80mm]{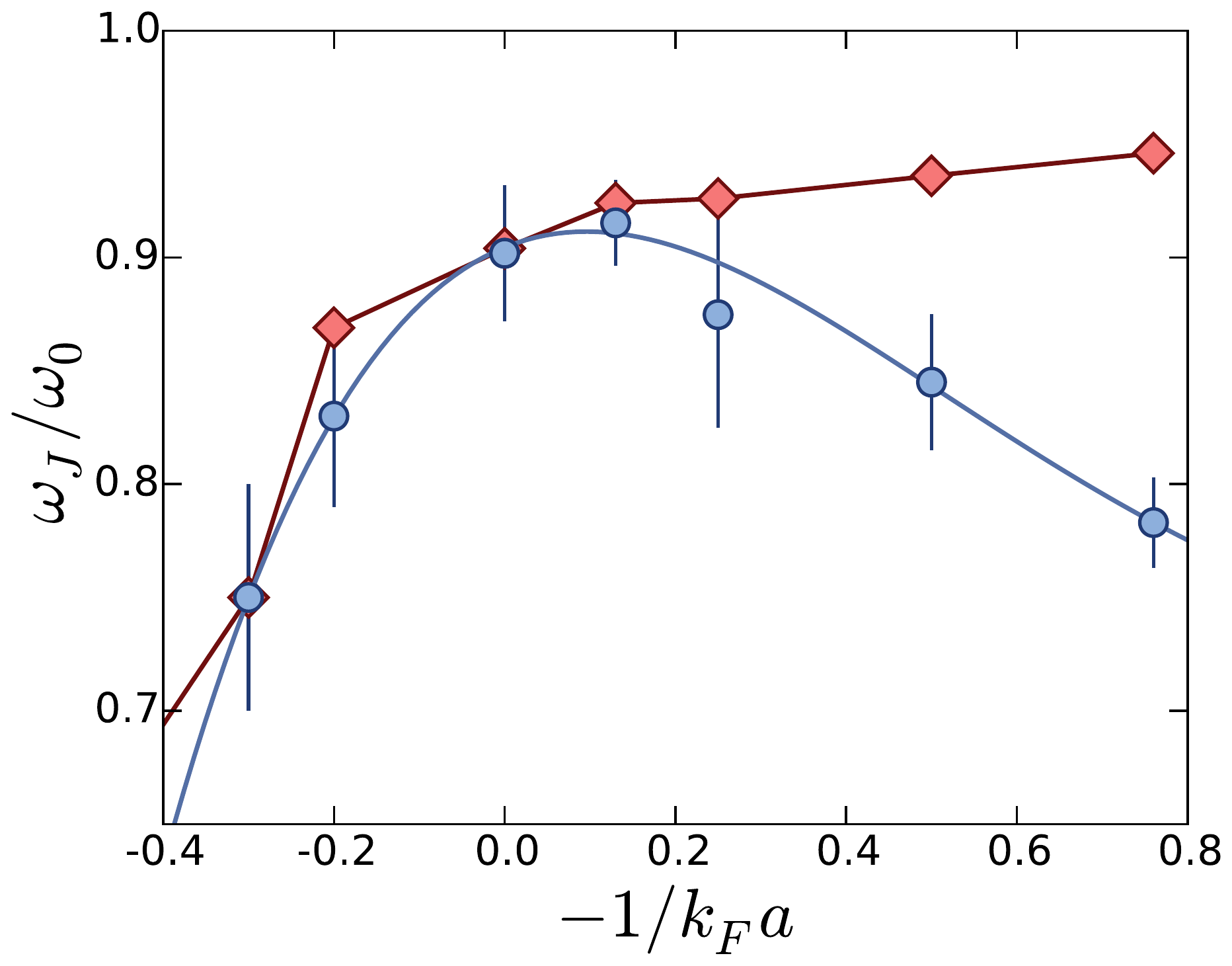}
\caption{Josephson frequency $\omega_J$ in units of the trap frequency $\omega_0$ for a fixed barrier height $V_0/E_F=1.2(1)$ as a function of 1/$k_\text{F}a$ around the resonance region: blue circles are the experimental findings, while red diamonds are the frequencies of the small oscillations calculated by numerically solving the time-dependent ETFM. The error bars in the experimental data are one standard deviation of five independent measurements.}
\label{FIG2S}
\end{figure}

The ETFM plasma frequency is obtained from the numerical solution of Eq.~(\ref{ETFM_eq}) with an initial imbalance $z(t=0) \equiv z_0 \to 0$: our results are plotted in Fig.~\ref{FIG2S} of this supporting material as red squares. We point out that, while the theoretical results are obtained in the limit $z_0 \to 0$, the experimental values for $\omega_J$ are taken extracting the frequency of population oscillation using the lowest detectable initial imbalance $z_0=0.03(1)$. However, the values of the plasma frequency predicted by the time-dependent ETFM are inside the error bars of the experimental data, from the BEC to the unitarity limit (see Fig.~\ref{FIG2S}). The agreement is eventually lost in the BCS side where the pair breaking effects on the tunneling are not correctly taken into account by the ETFM.
The charging energy $E_C$ can be calculated using the ETFM and it 
is given by $E_C=2{\cal U}/N$ where ${\cal U}=N\frac{\partial \mu_{loc}}{\partial N_L}$, the derivative being computed at $N_L=N/2$. $\mu_{loc}$ is the local chemical potential given by $\mu_{loc}=\int d\vec{r} \left[ \frac{\hbar^2}{2M} \left(\nabla \phi_\alpha\right)^2+V_p \phi_\alpha^2 + 
f(N_\alpha \phi_\alpha^2)\phi_\alpha^2\right]$ with $\alpha=L,R$ and 
$\phi_\alpha$ are the left and right Wannier wavefunctions computed 
from the ground and first excited states of the time-independent version of Eq. (\ref{ETFM_eq}) \cite{Sme03}. We studied the dependence of the chemical potential $\mu$ upon the barrier height verifying that the effect is of few percent (as expected since $\mu$ is a bulk property) and that an excellent estimate of ${\cal U}$ in our system is obtained by directly using $\mu$, computed with the barrier turned on, instead of using $\mu_{loc}$. The result of our study is plotted in Fig.~3B in the main text as a red diamonds. Typical values of $V_0$ for which the relative phase is well defined are $V_0 \gtrsim \mu$, with $\mu \sim 100 \hbar \omega_x$ on the BEC side ($1/k_F a \sim 4.5$) and $\mu \sim 500 \hbar \omega_x$ at the unitary limit. 
\begin{figure}[t!]
\centering
\includegraphics[width=80mm]{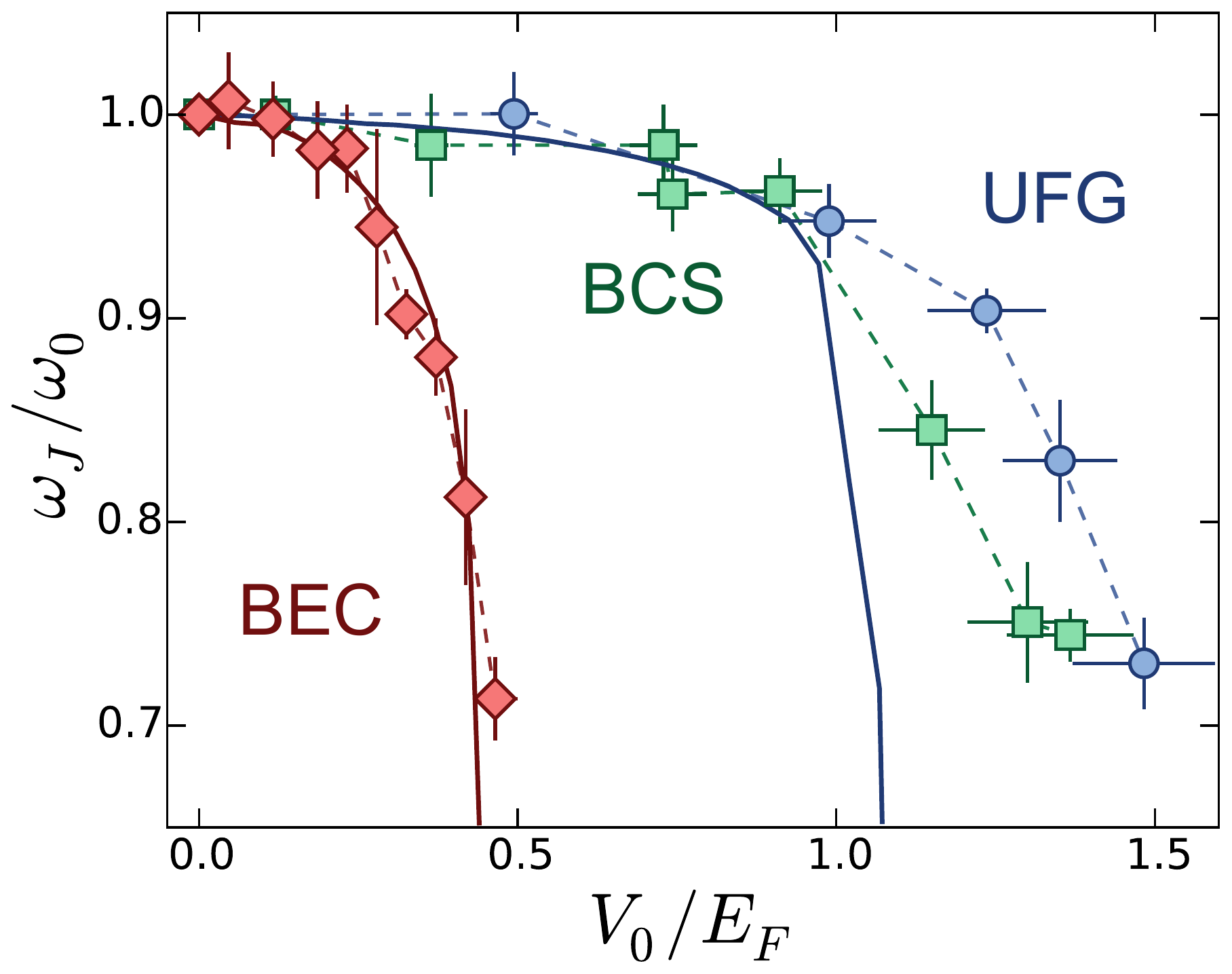}
\caption{Oscillation frequency $\omega_J$ in units of the trap frequency $\omega_0$ as a function of the normalized barrier height $V_0/E_{F}$, 
for a molecular BEC (red diamonds: $1/k_F a=4.6$), a unitary Fermi gas (blue circles: $1/k_F a=0$) and a BCS superfluid (green squares:$1/k_F a=-0.5$). The initial imbalance is $z_0=0.03(1)$. The corresponding values of ETFM predictions for the BEC (red solid curve) and unitary gas (blue solid curve) are also shown. In the numerical simulations the starting imbalance is set to $z_0=0.03$ in agreement with the experiment. The vertical error bars are one standard deviation of five independent measurements, while the horizontal error bars take into account the experimental uncertainties of the atoms number and of the barrier width.}
\label{FIG3S} 
\end{figure}
We investigated the values of $V_0$ for which the two-mode approximation \cite{Sme97a} gives a good description of the small oscillation frequency. In the two-mode model applied to Eq.~(\ref{ETFM_eq}), with the normalization of the wavefunction fixed to $N$ to have the correct chemical potential, the tunneling parameter ${\cal K}$ is the energy difference per particle between the first excited state and the ground state of the time-independent ETFM. By comparing the numerical results obtained from the numerical solution of Eq. (\ref{ETFM_eq}) with the two-mode model findings we found that a good agreement is present for $V_0\gtrsim1.4\mu$ in the BEC limit and $V_0\gtrsim1.2\mu$ at unitarity. In our system, we have $V_0\simeq1.1 \mu$ in the BEC side and we expect to have additional contributions beyond the two mode approximation. We also explored the system dynamics, using the ETFM, for larger starting imbalances and compared the simulation results with the experiment. In Fig.~\ref{FIG3S} we report the measured oscillation frequency for a molecular BEC ($1/k_F a=4.6$), a unitary Fermi gas and a BCS superfluid ($1/k_F a=-0.5$) as a function of the barrier height. In all cases the initial imbalance is $z_0=0.03(1)$. The theoretical oscillations frequencies, predicted by the ETFM, for the molecular BEC and the Unitary Fermi gas are also shown. In these simulations we fix the value of $z_0$ to the experimental one. We observe that, in all the explored interaction regimes, increasing $V_0$ leads to a progressive lowering of the oscillation frequency. As $V_0$ exceeds a critical value $V_0^{(cr)}$, the oscillations of $z(t)$ are eventually lost. In the BEC side one sees that the ETFM gives an excellent description also of the large oscillation regime. This is not the case for the unitary Fermi gas, where the ETFM severely underestimates $V_0^{(cr)}$, even though gives a good description of the small oscillations as seen in Fig.~\ref{FIG2S}.

\section{Observation and dynamics of topological defects}
\begin{figure}[htpb]
\centering
\includegraphics[width=90mm]{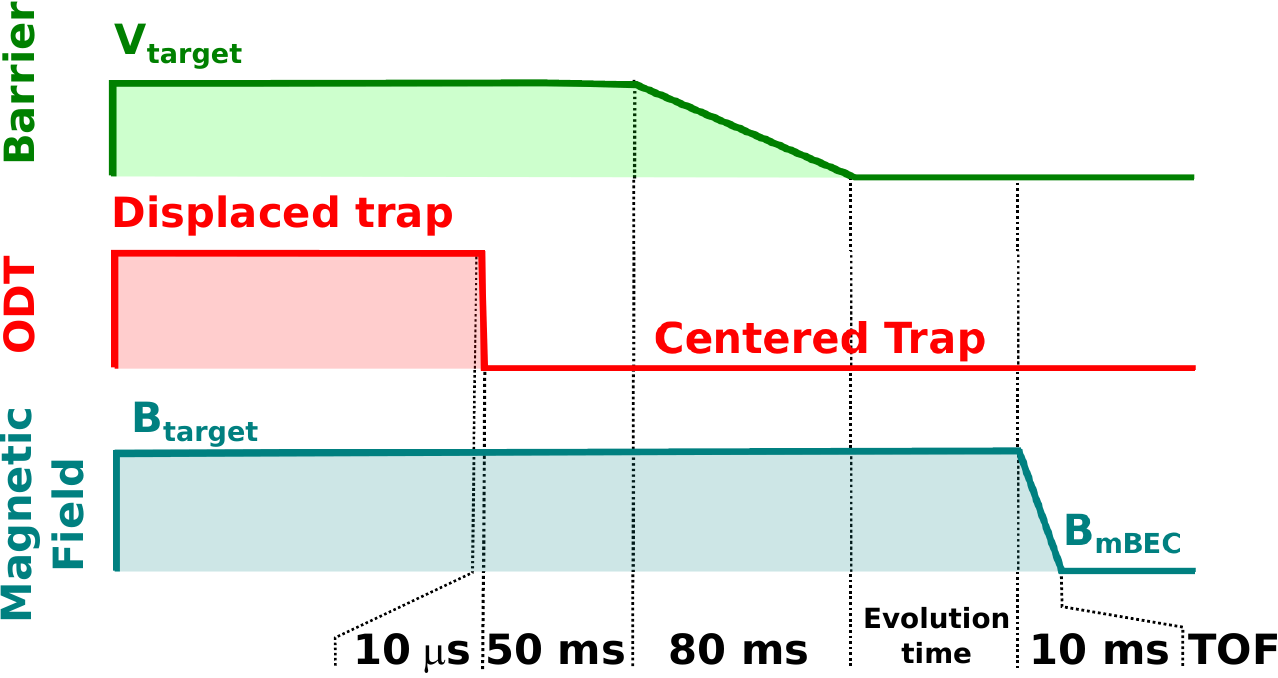}
\caption{Experimental procedure for detecting the topological defects.}
\label{FIG4S}
\end{figure}
At a critical value of the initial population imbalance and/or height of the  barrier, the system enters in a ``running phase" regime. This is accompanied by a continuous creation and annihilation of topological excitations inside the barrier region \cite{abadepl}. Depending on the geometry of the system and on the strength of the inter-particle interaction \cite{Piazza,recati}, the topological excitations can leave the low density region and escape inside the bulk of the system where they can be observed experimentally. The emergence of such dissipation mechanism is confirmed by the observation of vortices created at the center of the barrier and propagating in the direction of the superfluid flow. Their presence is revealed by following the procedure schematically illustrated in Fig.~\ref{FIG4S}. We initialize the tunneling dynamics and we let the system evolve for $50$ ms, a time scale longer than half of the axial trapping period. Successively the barrier is slowly switched off by linearly decreasing the laser power from the initial value to zero in $80$ ms. After a variable holding time, we turn off the optical confinement, and let the system expand for typically $10$~ms after which an image of the atomic density distribution is recorded. Once entering the running phase regime, we typically observe with non-zero probability a single defect, appearing as a density depletion in the cloud, see Fig.~4A in the main text. Before the time of flight expansion, the scattering length is tuned to the BEC side of the Feshbach resonance (690~G) by applying a $10$ ms linear ramp in the magnetic field. This procedure converts fermionic pairs into tightly bound molecules, emptying out the vortex core and enhancing the visibility of the topological defect \cite{Yefsah}.
 \begin{figure}[t]
\centering
\includegraphics[width=80mm]{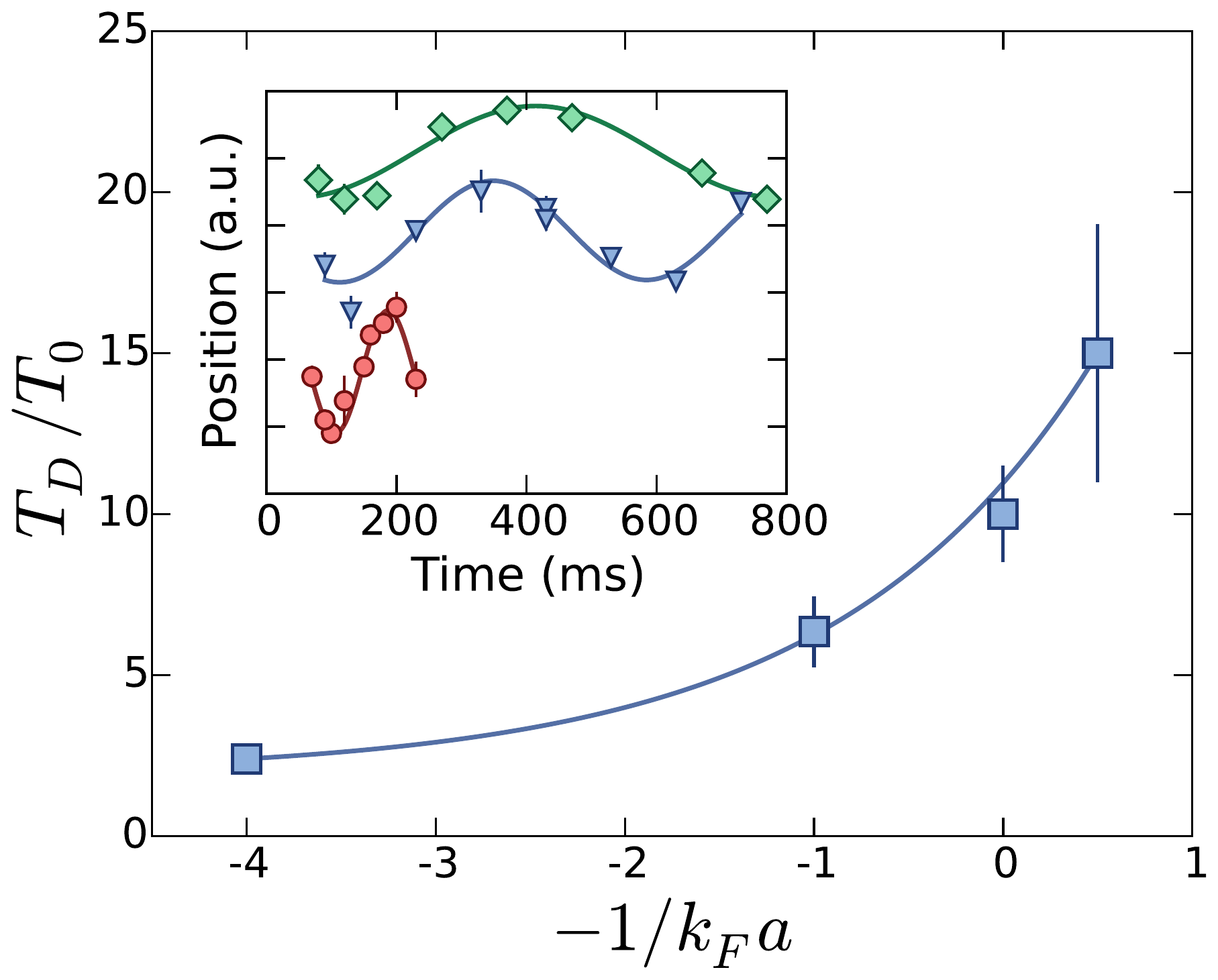}
\caption{Motion of topological defects. \textbf{Main:} Normalized period $T_{D}/T_{0}$ of a single defect as a function of $-1/k_Fa$ (blue squares), after removing the barrier. The blue line is a guide to the eye. The error bars correspond to the fit uncertainty. \textbf{Inset:} Axial position of a single defect versus time for $1/k_Fa$=4.25, 1, 0 (red circles, blue triangles, and green squares, respectively).   Here, the error bars are one standard deviation of five independent measurements in which the vortex was detected.}
\label{FIG5S}
\end{figure}
To gain more insights into the nature of the observed defects, we measure their axial oscillation period in the trap, once the barrier is removed. Fig.~\ref{FIG5S} shows the vortex period $T_{D}$, normalized to the trap period $T_{0}$=$2\pi/\omega_{0}$, as a function of $-1/k_Fa$. In the inset we instead present the evolution of the  position of the vortex in the trap for three different values of the interaction parameter. We note that the  vortex period  $T_{D}$ increases of about one order of magnitude moving from the BEC to the BCS side. This result is in agreement with the recent observation of solitonic vortices in BEC-BCS crossover superfluids, deterministically generated by phase-imprinting techniques \cite{Kua}. \\
We can infer that, in our system, the phase slippage process is induced by a vortex ring formed at the low density barrier region in a plane perpendicular to the flow \cite{Piazza}. However, this kind of defect is unstable when it moves in an anisotropic trap ($\omega_{y}$$\neq$$\omega_{z}$), as in our case, hence the ring will break into two vortex lines and finally will decay into a more stable structure which we can identify with a solitonic vortex in accordance with the large trap aspect ratio \cite{Komineas,Levin}. In future work, it would be interesting to further investigate the mechanism underlying the dissipation of the superfluid flow due to the propagation of topological defects as well as their dynamical evolution.

\end{document}